%% file: main.tex
\documentclass[letterpaper,twocolumn,10pt]{article}

\input{commands_packages}

\begin{document}

\title{Resolving Availability and Run-time Integrity Conflicts \\ in Real-Time Embedded Systems}

\author{Adam Ilyas Caulfield$^*$, Muhammad Wasif Kamran$^*$, N. Asokan$^{*,+}$ \\ $^*$University of Waterloo, $^+$KTH Royal Institute of Technology \\ {\tt acaulfield@uwaterloo.ca, mwkamran@uwaterloo.ca, asokan@acm.org}}

\maketitle

\begin{abstract}
Run-time integrity enforcement in real-time systems presents a fundamental conflict with availability.
Existing approaches in real-time systems primarily focus on minimizing the run-time overhead of monitoring. After a violation is detected, prior works face a trade-off: (1) prioritize availability and allow a compromised system to continue to ensure applications meet their deadlines, or (2) prioritize security by generating a fault to abort all execution. 

In this work, we propose \acron, an approach that offers a middle ground between the stark extremes of this trade-off. \acron monitors real-time tasks for run-time integrity violations and maintains an \textit{Availability Region (AR)} of all tasks that are \emph{safe to continue}. When a task causes a violation, \acron triggers a non-maskable interrupt to kill the task and continue executing a non-violating task within \AR. Thus, \acron ensures only violating tasks are prevented from execution, while granting availability to remaining tasks. With its hardware approach, \acron does not cause any run-time overhead to the executing tasks. \acron's design is formally verified, integrates with real-time operating systems (RTOSs), and is affordable to low-end microcontroller units (MCUs) by incurring +2.3\% overhead in memory and hardware usage.
\end{abstract}

\section{Introduction}
Embedded systems have become ubiquitous, often serving important roles in critical infrastructure responsible for real-time sensing or actuation, where deterministic and timely responses are crucial for functionality and safety. They are implemented with microcontroller units (MCUs), low-powered resource-constrained processing units optimized for energy efficiency. MCUs typically lack advanced features, such as memory management units (MMUs) or the ability to run rich operating systems (OS) that facilitate inter-process isolation. To make up for this, real-time operating systems (RTOSs) can be incorporated as an additional software layer to provide basic task management (e.g., scheduling, inter-task communication, timing services) with low overhead. Still, limited architectural protections make MCUs vulnerable~\cite{nafees2023smart}.

One prevalent threat includes run-time attacks, in which
an adversary (\adv) could inject code into memory, then exploit a vulnerability (e.g., a buffer overflow~\cite{cowan2000buffer}) to redirect execution to the injected code. 
Although memory-protection techniques can mitigate code-injection-based control flow attacks,
they do not protect against code-reuse attacks like return- or jump-oriented programming (ROP/JOP)~\cite{rop,jop}, that chain together carefully selected instruction gadgets from unmodified code. To account for such attacks, run-time integrity mechanisms are required. 

A prominent run-time integrity mechanism is Control Flow Integrity (CFI)~\cite{cfi-survey-1}, which employs reference monitors to detect when control flow instructions (e.g., calls, returns, jumps) deviate from an expected control flow path.
Recent CFI methods for embedded and real-time systems~\cite{cfi_rts_survey} enable cost-effective enforcement of valid control flows. Several designs~\cite{sherloc,recfish,insectacide,pac-pl,hcfi,eilid} focus on performing CFI checks in a way that minimizes run-time overhead on a program (or \textit{task}) while monitoring integrity. However, most works overlook the handling of tasks \textit{after} a CFI violation has been raised.
On the other hand, availability-focused security techniques for MCUs~\cite{garota,de2022casu,ecfi,aion,acfa,traces} do not account for run-time violations~\cite{garota,de2022casu,aion} or delay intervention until task completion~\cite{ecfi} or verification~\cite{acfa,traces,caramel}. 
Given this trade-off, it is unclear from current works how to achieve a middle ground between post-violation integrity and availability.

To bridge this gap, we introduce \acron: \textit{\textbf{P}reserving \textbf{A}vailabi- lity} \textit{and \textbf{I}ntegrity at \textbf{R}un-time}. \acron demonstrates that both availability and run-time integrity can be prioritized simultaneously. \acron hardware components monitor run-time integrity of independent tasks within a given program. Then, \acron dynamically adds and removes them from an Availability Region (\AR) based on their integrity status. When any run-time integrity violation is detected, \acron aborts execution of the violating task, removes it from \AR, and continues executing another task in \AR. \acron does not allow a task to return to \AR until a trusted software update occurs.
To summarize, we claim the following contributions:
\begin{myitemize}
    \item \acron, an architecture operating alongside an RTOS that maintains availability and integrity of benign tasks after integrity violations due to other tasks (Section~\ref{sec:overview});
    \item formally defined properties of \acron security, its modules sub-properties, and demonstration via model-checking that security goals are upheld (Sections~\ref{sec:models} and~\ref{sec:formal-sub-prop});
    \item design and implementation of \acron hardware submodules as formally-verified finite state machines (FSMs) that implement the outlined sub-properties (Section~\ref{sec:submodules});
    \item a FPGA prototype of \acron, evaluated with a real-world RTOS and example benchmark applications, demonstrating applicability to low-end MCUs due to minimal memory and hardware overheads (Section~\ref{sec:impl-eval});
    \item a response-time analysis of \acron's task revocation to understand its real-time implications by evaluating the conditions under which \acron's behavior maintains performance of non-violating tasks (Section~\ref{sec:rta});
    \item a case study of a simulated real-world real-time system executing atop \acron, showing that \acron effectively revokes access to violating tasks while maintaining core functionality of remaining tasks (Section~\ref{sec:case-study}).
\end{myitemize}

\section{Background}\label{sec:background}

\subsection{Control Flow Integrity (CFI)}
CFI is a countermeasure for control flow hijacking attacks, which exploit memory vulnerabilities to launch unintended run-time behavior by overwriting \textit{control data} (e.g., return addresses, indirect jump targets, function pointers). For example, an exploited buffer overflow can be used by \adv to perform ROP~\cite{rop}, in which \adv overwrites function return addresses to stitch together program gadgets to execute arbitrary (potentially malicious) behavior without corrupting instructions themselves and only corrupting return addresses. This can similarly be completed by corrupting \textit{forward-edge} control data, such as function pointers or indirect jump targets, to stitch together gadgets in the form of JOP~\cite{jop}.

CFI defends against such attacks by restricting the targets of control flow instructions (e.g., returns, calls, jumps) that use the control data that are exploited during control flow hijacking attacks, like ROP/JOP.
These instructions are restricted such that 
the target control data determining the destination of the control flow transfers is limited to a set of predetermined valid destinations.
Installing CFI into a system requires two components: (1) valid destination sets for each control flow instruction (at either coarse- or fine-granularity), and (2) a reference monitor to detect violations and respond accordingly~\cite{sok_cfa_cfi}. 


Many techniques employ their reference monitor through software instrumentation~\cite{cfi-care,ucfi,os-cfi,cfi-lb,urai,silhouette,vip-cfi,fineibt,typesqueezer,recfish,eilid}, enabling deployment of CFI without hardware changes.
In these works, software instrumentation either installs an in-line reference monitor which compares the control flow target to the valid set of destinations, or redirects to a trusted software module that implements the reference monitor. Despite being hardware-agnostic, many approaches based on software instrumentation assume the presence of an MMU and a rich OS to ensure the instrumentation is not tampered with at run-time. 

As these system mechanisms typically are not present on embedded systems, CFI mechanisms typically employ some type of hardware support, either in the form of custom hardware extensions~\cite{hafix,scfp} or by leveraging architectural features~\cite{silhouette,urai,sherloc,insectacide,cfi-care}. 
For example, features such as Branch Target Identification (BTI) and Pointer Authentication (PA) in ARM~\cite{arm_manual_bti_pac} have led to subsequent proposals for fully fledged CFI schemes~\cite{pac-ret,pacstack}. ARM has made BTI and PA available among their class of embedded devices, along with hardware tracing extensions that have also been leveraged to deploy CFI in embedded systems~\cite{sherloc,insectacide}.

Across particular approaches, one state-of-the-art approach for CFI is through implementing a shadow stack~\cite{burow2019sok}, a protected stack which is updated by the CFI reference monitor on each \texttt{call} instruction with the expected return address. Upon return, the CFI mechanism will compare the value that was popped from the normal stack to the expected value according to the shadow stack, and raise a violation when they do not match. This enables a fine-grained backward-edge defense, since each return address has exactly one target. Methods based on shadow stacks are effective across the device spectrum, and have been implemented in embedded systems~\cite{silhouette,flashadow,recfish}.

\subsection{Real-Time Operations in MCUs} 

For real-time operations, MCU software can use an RTOS for critical services (e.g., task scheduling, inter-task communication, timing adherence).
After an application assigns tasks priority and schedules them through an RTOS, the RTOS manages their execution. To do so, the RTOS contains functions to control execution, such as killing a task or yielding to another task. Tasks run until they voluntarily yield or until a time period has elapsed, at which point the RTOS performs a context switch and resumes the next highest-priority task.
Many works have proposed security mechanisms that account for real-time operations in embedded systems, including mechanisms to provide authenticated~\cite{iscflat,pearts,asap} and audited~\cite{acfa,traces,caramel} proofs of valid execution, availability mechanisms for trusted software~\cite{garota,aion,de2022casu}, and local run-time integrity in the form of CFI~\cite{cfi_rts_survey}.

CFI for real-time embedded systems is either fully software-based~\cite{ecfi,recfish,pac-pl} or uses hardware extensions with minimal software support~\cite{eilid,sherloc,insectacide}. Regardless of the approach, all aim to minimize disruptions to the executing task during the CFI monitoring. Upon the violation, they abort execution to prioritize run-time integrity, compromising the availability of a system in an unknown state. However, in the case of real-time systems, aborting execution system-wide can be costly.
An alternative approach addresses this trade-off by serving as a \textit{passive protection system} by logging the violation for later handling~\cite{ecfi}. However, this is equally unacceptable in this scenario, as it allows the hijacked task to continue with no guarantees regarding safety.

\subsection{Linear Temporal Logic}\label{sec:ltl}
We use Linear Temporal Logic (LTL)~\cite{ltl} to specify a formal model of \acron hardware and the properties that it should satisfy. As shown in prior works~\cite{garota,asap,vrased,apex,ucca}, LTL specifications can be used to describe finite state machines implemented within hardware and to reason about their properties. 

In particular, we define security properties and hardware submodules using LTL statements containing the following propositional connectives:
\begin{itemize}[nosep, leftmargin=*]
    \item conjunctions (e.g., logical \texttt{AND}) between two statements ($\land$); 
    \item disjunctions (e.g., logical \texttt{OR}) between two statements ($\lor$);
    \item and implications between statements ($\rightarrow$).
\end{itemize}
Additionally, we use the following temporal connectives between system states: 
\begin{itemize}[nosep, leftmargin=*]
    \item $X\phi$: denoting a \textit{next} state condition (e.g., condition $\phi$ is true in the next state); 
    \item $G\phi$: denoting a \textit{global} state condition (e.g., $\phi$ globally holds for all future states). 
\end{itemize}
In Section~\ref{sec:impl-eval}, we use model checking (NuSMV~\cite{nusmv}) to determine whether the system adheres to the specified properties. 

\subsection{Scope}\label{sec:scope}
We consider embedded systems operating real-time or sensing-based tasks atop low-end, single-core MCUs, with limited addressable memory (e.g., 8-256KB), split into program memory (PMEM) and data memory (DMEM). They are typically 8- or 16-bit CPUs running at 1-16MHz (e.g., TI MSP430). Such devices usually run software at \textit{bare metal}, meaning they lack MMUs, and thus perform memory accesses and execute instructions from  physical addresses. 

We consider the lowest end of MCUs, on which hardware features for access control~\cite{arm_mpu,msp430_ipe} are either limited or non-existent. For the sake of this work, we assume the latter. However, we do assume the MCU has an integrity monitor (e.g., CFI support) as this is widely studied and already deployed into real-world MCUs~\cite{arm_manual_bti_pac}. We consider implications of more advanced platforms in Section~\ref{sec:discussion}.

In terms of software, we assume modularized set of  \textit{tasks} execute on the MCU. Each task has differences in timing requirements and priority.
To manage these task characteristics, we assume an RTOS is also in use for scheduling, timing services, and synchronization between the tasks.

\section{Problem Statement}\label{sec:problem-statement}

\subsection{Motivating Example}\label{sec:application}

Consider a simple MCU that is used to implement a syringe pump controller that allows physicians to monitor and deliver medication to patients remotely over the network~\cite{islam2019design,saqlain2023iot}. In this scenario, the device may be installed with multiple software submodules serving different roles, operating as separate tasks atop an RTOS for prioritization, scheduling, and timely execution. For example, it may include the following:
\begin{myitemize}
    \item a WiFi module for network communication,
    \item LCD display and buzzer control for user interface and feedback,
    \item sensor monitoring of patient vitals (e.g., temperature and heart-rate monitoring), and
    \item motor control for accurate and timely dosage injection.
\end{myitemize}

Suppose the network communication task contains a buffer overflow~\cite{cowan2000buffer} on the stack, enabling a control flow hijack attack (e.g., ROP~\cite{rop}). Given a CFI integrity monitor has been installed in the device, attempts to exploit this vulnerability will be detected. As discussed in Section~\ref{sec:background}, current approaches would not offer a suitable balance: they would either suspending all tasks for the duration of the reset and recovery window, including motor control and sensor monitoring, or allow the compromised network communication module to continue without safety guarantees.

Scenarios like this demand a mechanism that can simultaneously prevent violating tasks while ensuring non-violating tasks can continue their time-sensitive execution.
Critically, hardware-based mechanisms are required, as other software (e.g., tasks or the RTOS itself) may attempt to reschedule the violating task before it is properly remediated. We revisit this example as a case study in Section~\ref{sec:case-study}.

\subsection{\acron at a High Level}\label{sec:overview}

Unlike a typical system in which a run-time integrity monitor (\IM) would directly invoke a system reset signal, \acron's goal is to ensure non-violating tasks can continue safely after a run-time integrity violation has been detected. Therefore, upon receiving a redirected violation signal, \acron can perform operations to ensure benign tasks can securely continue.
\acron goals can be reduced to upholding the following two security properties at run-time:
    
\noindent \textbf{[P1] Post-Violation Availability:} when a task has a run-time integrity violation, \acron always (1) removes the task from \AR and (2) triggers the yielding to a non-violating task;
    
\noindent \textbf{[P2] Post-Violation Integrity:} tasks outside \AR can never re-execute until a software update has occurred. 

\begin{figure}[t]
    \centering
    \includegraphics[width=\columnwidth]{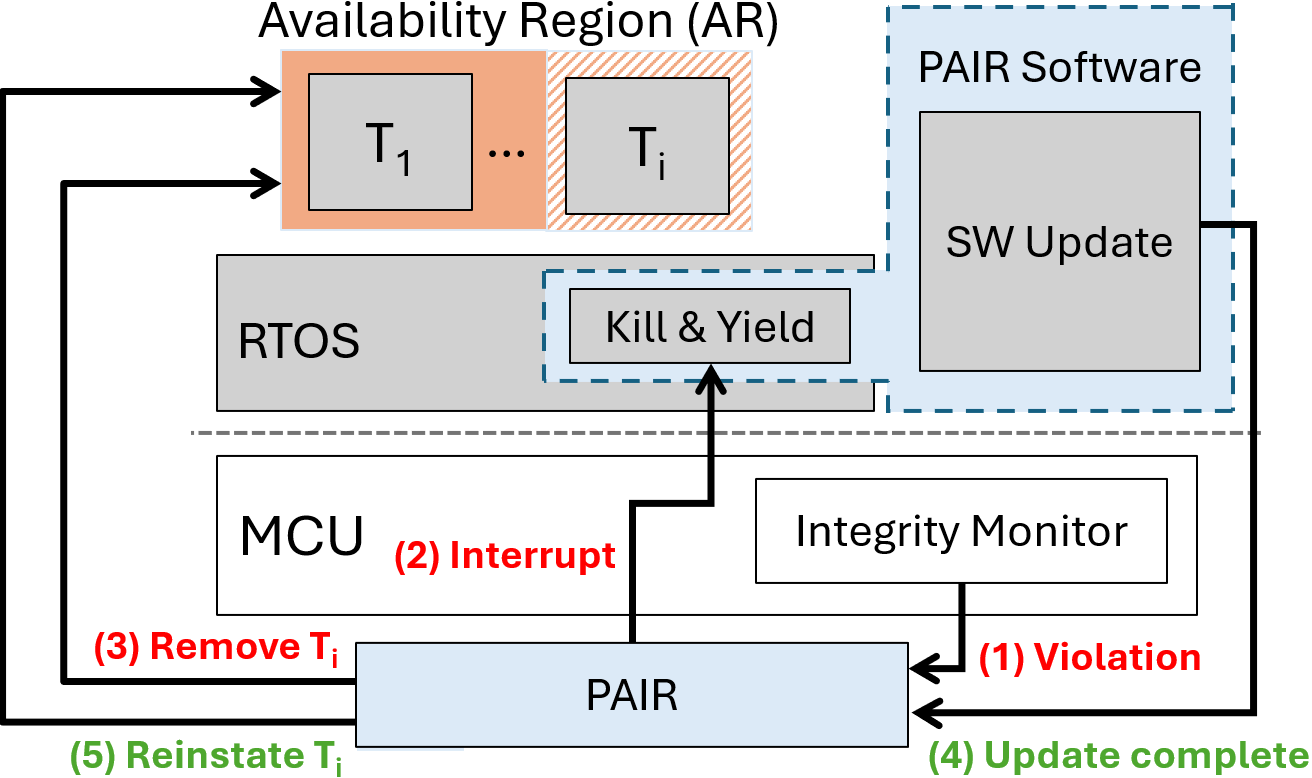}
    \caption{\acron Overview: Tasks and an RTOS execute in software, with all tasks initially in the \AR. After detecting an integrity violation (1), \acron triggers the trampoline into RTOS to \textit{kill-and-yield} (2) and removes the violating task from \AR (3). Only after a trusted software update has completed (4) will \acron reinstate the violating task into \AR (5).}
    \label{fig:system}
\end{figure}

A high-level overview of an \acron-enabled system is depicted in Figure~\ref{fig:system}.
It assumes a system that operates real-time tasks (denoted $T_{i}$) with an RTOS, while \acron introduces two trusted software components. 
Initially, the MCU runs with no violations, and \acron has assigned all tasks to \AR.
\acron monitors MCU signals and memory to determine which real-time task is executing at any given moment. \acron interfaces with an \IM to detect the violation when it occurs (1). Upon this detection, \acron will generate a non-maskable interrupt (2) to invoke the \textit{kill-and-yield} trampoline, while also removing the violating task from \AR. These steps are repeated for any other tasks that generate a violation. The bounds of \AR are not expanded until a trusted software update has completed (4), and at this point \acron will reinstate all violating tasks back into \AR.

\section{Adversary \& System Models}\label{sec:models}

\subsection{Adversary Model}\label{sec:adv}
We consider \adv that exploits software vulnerabilities to manipulate any data or software (tasks or the RTOS), unless it is explicitly protected by hardware.
\adv aims to cause run-time integrity violations (e.g., control flow hijack and ROP/JOP attacks~\cite{jop,rop}) to invoke malicious actions or avoid critical actions.
\adv aims to continue its malicious actions even after detection by re-entering revoked tasks. 
Physical attacks to circumvent hardware protections are out of scope for this work, as they require orthogonal physical access control measures~\cite{obermaier2018past}. 

\subsection{System Model}

\begin{figure}[t]
    \centering
    \includegraphics[width=0.95\linewidth]{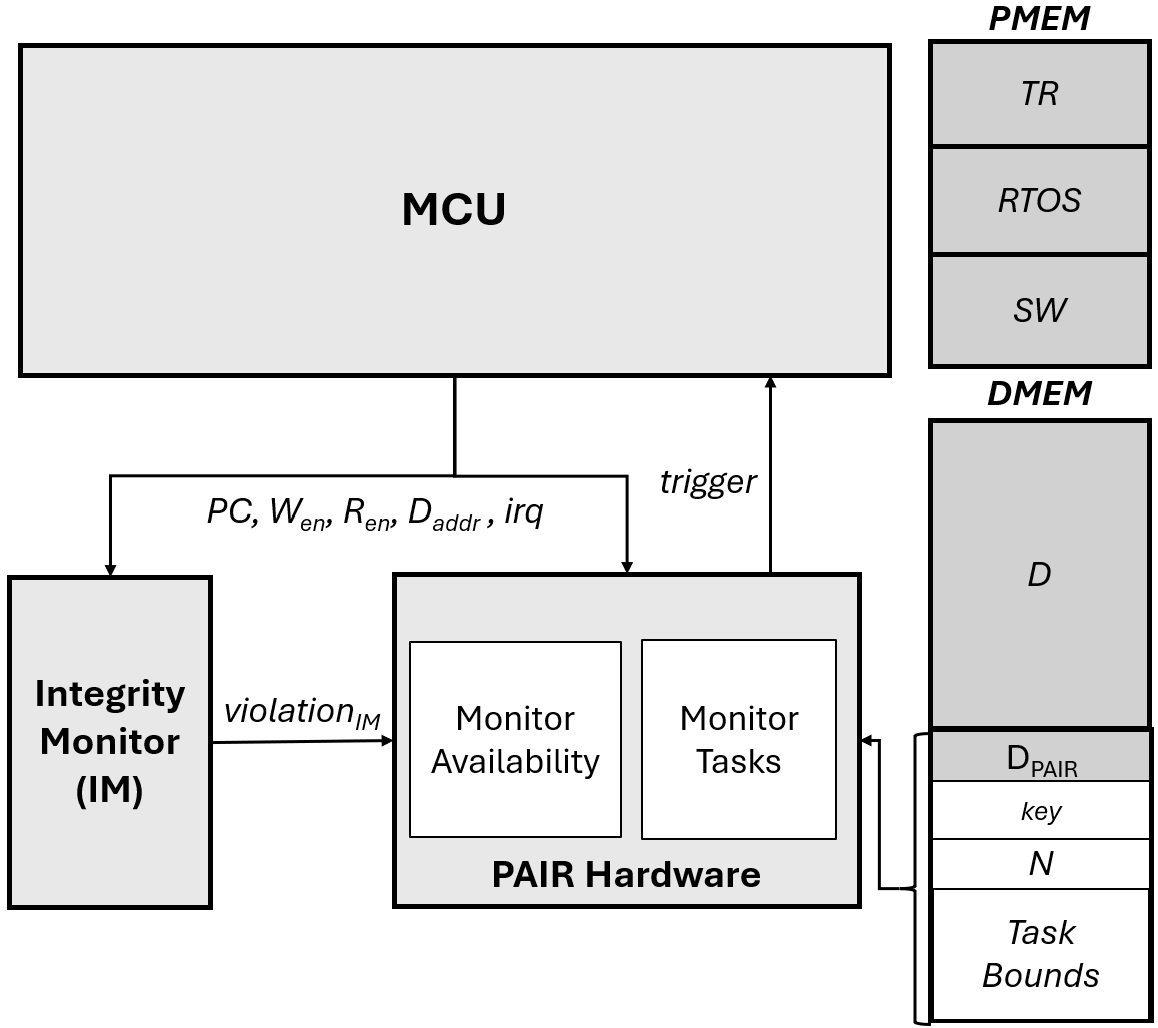}
    \vspace{-1em}
    \caption{System Configuration: \acron interfaces with MCU, an Integrity Monitor (\IM), and memory}
    \label{fig:system-config}
\end{figure}

Figure~\ref{fig:system-config} shows the assumed initial hardware configuration and memory layout.
\acron interfaces with the MCU to read the following signals pertaining to its run-time behavior:
\begin{myitemize}
    \item \textit{PC}: the program counter register storing the memory address of the instruction that is executing;
    \item $W_{en}$: a flag denoting whether the instruction is performing a write to memory;
    \item $R_{en}$: a flag denoting whether the instruction is performing a read from memory;
    \item $D_{addr}$: the memory address targeted by the current instruction (i.e., the address being written/read if $W_{en}$/$R_{en}$);
    \item \textit{irq}: MCU signal that indicates whether an interrupt is happening (i.e., the MCU has identified the interrupt source and is currently interrupting the current execution).
\end{myitemize}
We assume the MCU is extended with an existing \IM that interfaces with relevant MCU signals. We assume the \IM outputs a signal \textit{$violation_{IM}$} that is only set when the current instruction is violating its security policy. This signal is passed into \acron hardware.

MCU memory is divided into program memory (PMEM) and data memory (DMEM). PMEM contains fixed sub-regions for the following components:
\begin{myitemize}
    \item \TR: the task region containing all real-time tasks;
    \item \textsf{RTOS}: the region containing the RTOS itself;
    \item \sw: the region containing \acron trusted software.
\end{myitemize}
DMEM contains the following fixed sub-regions:
\begin{myitemize}
    \item $\mathsf{D}$: memory used for tasks and RTOS;
    \item $\mathsf{D_{\acron}}$: all critical data referenced by \textsf{SW} or hardware.
\end{myitemize}
$\mathsf{D_{\acron}}$ contains a \textit{key} used by \textsf{SW} trusted update function, the number of tasks configured in \TR ($N$), and their bounds within \TR (i.e., for each $T_{i}$, the task bounds are specified as a $T^{min}_{i}$ and $T^{max}_{i}$ holding the minimum and maximum PMEM addresses pertaining to $T_{i}$).
We assume that task bounds are non-overlapping and independent in terms of their code, but may share data.
\acron hardware monitors signals from MCU and \IM to monitor tasks and their availability (i.e., maintain \AR), and outputs a \textit{trigger}, which implements a non-maskable interrupt to invoke \textsf{SW}. 

\subsection{Formalized Security Properties}\label{sec:formal-goals}
We formalize the security properties of \acron in Figure~\ref{fig:formal-goals}. We define \textbf{[P1]} as the two LTL statements in Definition 1.
Given a run-time integrity violation is occurring (\textit{$violation_{IM}$} is true) or another access-control violation to \acron critical components (\textit{$violation_{\acron}$} is true), \acron's kill-and-yield trigger (\textit{trigger}) should be set in the next state. 
Given task $i$ ($T_{i}$) is executing (i.e., $PC$ is within the bounds of \TR and the current $task_{id}$ is $i$) and \acron has generated \textit{trigger}, the violating task $T_{i}$ should no longer be in \AR in the next state.
We define \textbf{[P2]} in Definition 2. If $T_{i}$ is executing, it must also be true that either $T_{i}$ is in \AR or that \textit{trigger} will be set in the next state.

\section{Security Sub-properties and Proofs}\label{sec:formal-sub-prop}

\subsection{\acron module Sub-properties}
Based on \acron's goals, we outline a set of required sub-properties to be implemented by \acron's design. These sub-properties fall into two classes: task monitoring and availability monitoring.
Later in Section~\ref{sec:formal-proof}, we demonstrate that these sub-properties suffice to achieve \acron's desired security properties.

An axiom as a part of the task monitoring specification is the interaction with $\mathsf{D_{\acron}}$ to read the task boundaries. This is outlined in Figure~\ref{fig:task-mon}. Given a fixed set of tasks $N$ with their bounds installed in fixed locations, we assume LTL~\ref{ltl:track-task} that specifies the configuration of an internal label $task_{id}$. This label is set depending on which task bounds $PC$ is currently within. 
Given $PC$ is within the bounds of $T_{i}$, $task_{id}$ is set to $i$. 

\input{ltl_specs/goals_ltl}
\input{ltl_specs/task_monitor_ltl}
\input{ltl_specs/avail_mon_ltl}

Two additional sub-properties are required to ensure that tasks do not perform any illegal memory accesses. These sub-properties are also outlined in Figure~\ref{fig:task-mon}.
The sub-property specified in LTL~\ref{ltl:task-mon-violation-2} requires that \acron trigger \textsf{SW} to remove task $i$ from \AR when that task attempts to read or write to $D_{\acron}$ while outside of \textsf{SW}. This is defined by setting \textit{$violation_{\acron}$} when a read or write instruction is targeting $D_{\acron}$ while not executing within \textsf{SW}.
The second sub-property specifies that \textit{$violation_{\acron}$} should be set when software that is not in \textsf{SW} attempts to modify PMEM, as defined in LTL~\ref{ltl:task-mon-violation-3}. Whenever \textit{$violation_{\acron}$} is set, \textit{trigger} must be set in the next state, as specified in LTL~\ref{ltl:pair-violation}.

\textit{\textbf{\underline{Note:}} Further access control is possible if implemented in the \IM. For example, our prototype \IM (described in Section~\ref{sec:impl-eval}) enforces CFI and prevents tasks from modifying each other's stacks.}

The required sub-properties for monitoring availability and maintaining \AR are outlined in Figure~\ref{fig:aviail-mon-spec}.
\acron maintains a bit vector $AR_{en}$ that signifies whether a task is a part of the \AR: removing or adding a task to \AR corresponds to clearing or setting its corresponding bit in $AR_{en}$.
Therefore, changes to \AR are formally specified by bitwise operations that clear or set corresponding bits in $AR_{en}$. To model this, we define \texttt{mask}($id$) as a method to map an $id$ to an $N$-length bitmask. Similarly, we define \textit{zero} and \textit{set} as $N$-length bitmasks of all zeros and ones, respectively.

LTL~\ref{ltl:generate-irq} and~\ref{ltl:clear-irq} specify the requirements to control \acron's non-maskable interrupt signal \textit{trigger}.
LTL~\ref{ltl:generate-irq} shows that \textit{trigger} should be set in the next state when \textit{$violation_{IM}$} has been set by the \IM. 
LTL~\ref{ltl:clear-irq} shows that \textit{trigger} is cleared in the next state when \textit{trigger} is currently set and the \textit{irq} has been set, denoting that the interrupt has been accepted by the MCU. 

LTL~\ref{ltl:update-AR-trigger} requires that the bit in $AR_{en}$ corresponding to the executing task is cleared when \textit{trigger} has been generated. This is specified by requiring that the bitwise AND of \texttt{mask}($task_{id}$) and $AR_{en}$ in the next state equals zero. 

Any attempts to re-execute a task that has been removed from \AR should also result in the invocation of \acron's software to yield to another non-violating task. This is specified by LTL~\ref{ltl:trigger-on-reenter} by requiring that \textit{trigger} is set in the next state when the bit in $AR_{en}$ corresponding to $T_{i}$ has been cleared. Finally, $\AR$ should only be reset (i.e., all bits in $AR_{en}$ set to one) when exiting from \textsf{SW}. This is specified in LTL~\ref{ltl:reset-AR}.

\subsection{\acron goals via sub-properties}\label{sec:formal-proof}
Figure~\ref{fig:proofs} outlines the theorems that state properties \textbf{[P1]} and \textbf{[P2]} are upheld. 
These theorems are also checked via model checker, described further in Section~\ref{sec:impl-eval}.
Here we also discuss the intuition behind why they are upheld.

\input{ltl_specs/proofs}

Theorem 1 states that LTLs~\ref{ltl:task-mon-violation-2},~\ref{ltl:task-mon-violation-3},~\ref{ltl:pair-violation},~\ref{ltl:generate-irq}, and~\ref{ltl:clear-irq} uphold Definition 1 for \textbf{[P1]}. Definition 1 specifies that each instance of \textit{$violation_{IM}$} or \textit{$violation_{\acron}$} must always result in \textit{trigger} in the next state. It is provided by LTLs~\ref{ltl:generate-irq} and~\ref{ltl:clear-irq} that \textit{trigger} is set at this instance and is not cleared until the MCU generates its interrupt signal, denoting it was accepted. Additionally, LTLs~\ref{ltl:task-mon-violation-2} and~\ref{ltl:task-mon-violation-3} ensure that a task's memory accesses relevant to \acron result in \textit{trigger}, regardless of the specific \IM in use.

Theorem 2 states that LTLs~\ref{ltl:track-task},~\ref{ltl:update-AR-trigger},~\ref{ltl:trigger-on-reenter},~\ref{ltl:reset-AR} uphold Definition 2 for \textbf{[P2]}.
Definition 2 describes post-violation integrity by specifying that execution of a task in a given state implies either that the task is in \AR or that \textit{trigger} will be generated in the next state.
This property requires monitoring of the currently executing task, which is specified in LTL~\ref{ltl:track-task} to be maintained in $task_{id}$. LTL~\ref{ltl:update-AR-trigger} specifies that given \textit{trigger} has been generated, the current task is removed from \AR by performing the corresponding bitwise operations based on $task_{id}$. Additionally, LTL~\ref{ltl:trigger-on-reenter} supports Definition 2 by ensuring any attempted re-entering of a removed task results in \textit{trigger} generation, specified by generating \textit{trigger} when \textit{$AR_{en}$} shows the current task was revoked from \AR.
Finally, LTL~\ref{ltl:reset-AR} also supports Definition 2 by specifying tasks can be re-added to \AR after exiting from \textsf{SW}, and thus they will not generate \textit{trigger} if they have been re-executed after a software update.

\section{Implementation \& Verification}\label{sec:submodules}
\acron modules are implemented as two finite state machines (FSMs) controlling \textit{trigger}=and \textit{$AR_{en}$}, respectively. These are depicted in Figures~\ref{fig:fsm-trigger} and~\ref{fig:fsm-ar}.

\subsection{Sub-module FSMs}

Figure~\ref{fig:fsm-trigger} shows the FSM for handling \textit{trigger}. The value of \textit{trigger} is set based on either of the following states:
\begin{myitemize}
    \item \textit{Exec}: representing the state in which the MCU is executing without a violation, hence \textit{trigger} is unset;
    \item \textit{Trigger}: representing the state in which a violation has occurred and \textit{trigger} is set.
\end{myitemize}
The MCU enters \textit{Exec} upon any reset, and thus this is the initial state. A transition from \textit{Trigger} into \textit{Exec} occurs based on the specification in LTL statements~\ref{ltl:task-mon-violation-2},~\ref{ltl:task-mon-violation-3}, and~\ref{ltl:generate-irq} corresponding to run-time violations. A transition from \textit{Trigger} into \textit{Exec} only occurs when the \textit{trigger}-based interrupt has been accepted by the MCU, corresponding to LTL statement~\ref{ltl:clear-irq}.

\input{figs/fsm-trigger}

Figure~\ref{fig:fsm-ar} shows the FSM for modifying \AR via the bitvector \textit{$AR_{en}$}. Each state corresponds to the status of $AR_{en}$:
\begin{myitemize}
    \item \textit{Monitor}: state corresponding to execution that in which tasks are monitored. This state does not require reinstating or revoking a task to/from \AR; thus, $AR_{en}$ remains unchanged in this state; 
    \item \textit{Revoke}: state corresponding to that in which a task must be removed from \AR, and the corresponding bit of the violating task is cleared in this state; 
    \item \textit{Reinstate}: state corresponds to a successful software update and thus all tasks are reinstated in \AR.
\end{myitemize}
This FSM initializes in the \textit{Monitor} state. Only upon a trigger does a transition into \textit{Revoke} occur, and
a transition from \textit{Monitor} into \textit{Reinstate} occur
only occurs when a software update completes (i.e., $PC$ being equal to $SW_{exit}$). Both \textit{Revoke} and \textit{Reinstate} states only execute for one cycle to update \AR accordingly before transitioning to \textit{Monitor}. The transition from \textit{Monitor} into \textit{Revoke} corresponds to LTL~\ref{ltl:update-AR-trigger}, and the transition from \textit{Monitor} into \textit{Reinstate} corresponds to LTL~\ref{ltl:reset-AR}.

\input{figs/fsm-ar}

\subsection{Formal Verification}
We use NuSMV~\cite{nusmv} as a model checker to determine whether \acron design adheres to the specified properties and statements from Section~\ref{sec:formal-sub-prop} are upheld. We use Verilog2SMV~\cite{verilog2smv} to convert Verilog specifications into nuSMV statements. Then, we formulate the LTL specifications in nuSMV to check the model. We execute the model checking atop an Intel(R) Core(TM) Ultra 7 155H (3.80 GHz) with 32.0 GB RAM using Ubuntu-20.04 under Windows Subsystem for Linux 2 (WSL 2). The model checker executed for 43.98 seconds with peak memory usage of 1.38GB to complete the verification.

\section{Prototype \& Evaluation}\label{sec:impl-eval}

\subsection{Prototype configuration}\label{sec:impl}

To implement \acron prototype,
we use the Xilinx Vivado tool-set~\cite{vivado}
as a development platform. \acron hardware is written in the Verilog hardware description language and implements each of \acron sub-modules according to the logic defined in Section~\ref{sec:submodules}, and its software is implemented in C. 
Using Vivado, we synthesize \acron hardware, simulate behavior, evaluate area/energy costs, and deploy a \acron-equipped openMSP430~\cite{openmsp430}
on the Basys3 prototyping board~\cite{basys3}
with 16KB of PMEM and 12KB of DMEM.

A custom linker is used to ensure all code and data of interest (from Figure~\ref{fig:system-config}) are placed into contiguous memory regions that are monitored by \acron. 
For the RTOS in our evaluation, we use RIOT~\cite{riot}
due to its open-source availability, support for MCUs within our scope, and use in prior works~\cite{aion}. In theory, \acron concepts can apply to any RTOS with an identifiable API for \textit{kill} and \textit{yield} procedures.

For the \IM, we instantiate a task-aware shadow stack module along with an indirect-branch table (IBT) in hardware for CFI of return and jump instructions, respectively. Both interface with memory and MCU signals as required.
Additionally, it monitors tasks' stacks to ensure they do not interfere with each other's temporary variables.

\input{figs/hw-cost}

\subsection{Fixed Costs}
\acron hardware cost is accounted for in the Look-up tables (LUTs) and flip-flop registers (FFs) added to the openMSP430 core. We compare this cost to that of closely related works in Figure~\ref{fig:hw-cost}.
\acron requires fewer resources than all closely related works that deploy security extensions atop openMSP430.
Although \acron has a different goal than each of these works, this shows that it is cost-effective for low-end devices and is in line with related works that target the same class of devices.
The most closely related works are EILID~\cite{eilid} and AION~\cite{aion}. EILID implements CFI for low-end MCUs atop a hybrid (software-hardware) root of trust CASU~\cite{de2022casu} for memory immutability and trusted software updates. 
We note that when including \IM in the cost, \acron + \IM requires more hardware resources than EILID.
However, we note that \acron is concerned with providing availability and run-time integrity simultaneously, whereas EILID's primary focus is run-time integrity (discussed further in Section~\ref{sec:related-work}).
AION builds atop Sancus~\cite{Sancus17} hardware-based TEE, which requires a more extensive set of security features.
We also evaluate \acron performance overheads by considering additional energy consumption. To evaluate energy overheads, we use Vivado synthesis tools to estimate \acron's power consumption on the Basys3 FPGA board.
We find the openMSP430 core without \acron consumes 71 mW of static power and 76 mW with \acron, showing that \acron incurs only an additional 5 mW in static power consumption. 
The dynamic power consumption of the simulated MCU board was measured at 109 mW. The contribution from PAIR was negligible, failing below Vivado’s minimum reportable threshold of 1 mW.  

\subsection{Performance costs}\label{sec:performance-costs}

We evaluate the performance costs of \acron using the run-time and memory overheads when executing software atop \acron.
\acron hardware does not impose any run-time overheads, since it is monitoring MCU signals without affecting the MCU's critical path. 
The memory cost incurred by \acron varies per application. For instance, $\mathsf{D_{\acron}}$ size depends on the number of independent tasks. Memory cost may also vary based on the particular \IM in use.
Our \IM instance memory usage for shadow stacks depends on the maximum call depth among all tasks, and IBT size depends on the total number of indirect forward edge control flow transfers among all tasks. To effectively evaluate \acron memory usage, we select a set of applications from RIOT test examples and BEEBs embedded systems benchmark suite~\cite{beebs_repo}. We select RIOT test examples that create multiple tasks, and BEEBS applications that meet one or more of the following criteria: 
\begin{enumerate}[nosep=*]
    \item has one or more recursive calls;
    \item has one or more indirect forward branches (jump/call);
    \item has \textit{high} or \textit{very high} suitability for MCUs~\cite{beebs_repo}.
\end{enumerate}
Table~\ref{tab:memory_usage} shows the resulting memory usage.

\input{figs/memory_usage}

Among the selected applications, \acron memory usage is minimal, an increase ranging from just 4 to 16 bytes (0.024\% to 0.097\% of available DMEM).
Additional memory usage is required due to the \IM, but it is still minimal: an increase ranging from 8 to 42 bytes (0.048\% to 2.28\% of the available data memory). \acron by itself has a minimal fixed memory cost ($2 + 2\times N$ bytes), but incurs additional variable costs due to the \IM that is installed.

Finally, we evaluate the latency overhead introduced by \acron's interrupt path relative to the native software call of the \textit{kill-and-yield} routine. We measure this latency across four scenarios, varying two factors. First, we vary the direction of the priority switch from high-to-low (H$\rightarrow$L) and from low-to-high (L$\rightarrow$H). Second, we vary the invocation source between the native call vs. PAIR-triggered interrupt, the handler of which jumps to the same function. 

\input{figs/irq-latency}

As shown in Figure~\ref{fig:kill-and-yield-time}, \acron interrupt path introduces a consistent overhead of approximately 3.5$\mu$s relative to the native call path across both priority directions (136.7 $\mu$s vs. 133.2$\mu$s for H$\rightarrow$L and 134.7$\mu$s vs. 131.0$\mu$s for L$\rightarrow$H), reflecting the cost of hardware violation detection and non-maskable interrupt generation. The H$\rightarrow$L transition incurs marginally higher latency in both cases, due the scheduler identifying the next task. The negligible and consistent times confirm that \acron introduces minimal overhead.

\section{Response Time Analysis (RTA)}\label{sec:rta}

Terminating a violating task frees system execution time, reducing the response time of remaining lower-priority tasks, but this benefit may be offset by \acron's interrupt-path latency  (see Section~\ref{sec:performance-costs}). To analyze this tension, we apply fixed-priority RTA: we first establish a baseline RTA model, extend it to capture \acron's revocation behavior and measured interrupt-path cost, then derive the condition under which interference from a higher-priority task eliminated by revocation offsets this enforcement cost. Table~\ref{table:rta-notation} summarizes notation.

\subsection{Baseline RTA Model}

We first establish a baseline using classical fixed-priority RTA~\cite{joseph1986finding}.
For a task $i$, the worst-case response time is given by the least non-negative fixed point of Equation~\ref{eq:template-resp-time}:

\begin{equation}\label{eq:template-resp-time}
R^{(n)}_i =
C_i + B_i +
\sum_{j \in hp(i)}
\left\lceil
\frac{R^{(n-1)}_i}{T_j}
\right\rceil C_j,
\end{equation}

In this equation, $C_i$ and $B_i$ denote the worst-case execution time and blocking time of task $i$, respectively. The terms $C_j$ and $T_j$ denote the worst-case execution time and period of higher priority task $j$. 
The set $hp(i)$ denotes all tasks with higher priority than task $i$. 
For each task $j$ in $hp(i)$, the ceiling term $\lceil R^{(n-1)}_i / T_j \rceil$ gives the number of times task $j$ can execute within task $i$'s response time interval as estimated at the previous iteration. It is multiplied by $C_j$ to give the task's total interference contribution. The interference contributions of all higher priority tasks are summed to obtain the total interference, which is summed with $C_i$ and $B_i$ to obtain the worst-case response time. 

\begin{table}[t]
    \centering
    \renewcommand{\arraystretch}{1.4} 
    \caption{Response Time Analysis Notation}\label{table:rta-notation}
    \vspace{-1em}
    \footnotesize{
    \begin{tabular}{|c|p{0.77\columnwidth}|}
         \hline
         \textbf{Notation} & \textbf{Meaning} \\
         \hline
         $C_i$, $C_k$, $C_l$ & the WCET of tasks $i$, $k$, and $l$, respectively \\
         \hline
         $B_i$ & the blocking time of task $i$ \\
         \hline
         $hp(i)$ & the set of tasks that are higher priority than task $i$ \\
         \hline
         $T_{i}$, $T_{k}$, $T_{l}$ & the period of tasks $i$, $k$, and $l$, respectively \\
         \hline
         $R^{(n)}_i$ & the worst-case response time of task $i$ computed at iteration $n$ \\
         \hline
         $R_i$ & the worst-case response time of task $i$ after convergence \\
         \hline
         $\mathit{budget}_k$ & the amount of execution time consumed by task $k$ before \acron revokes it \\
         \hline
         $R^\text{baseline}_i$ & the converged worst-case response time of task $i$ in the baseline model \\
         \hline
         $n_k$, $n_l$ & simplified ceiling terms of task $k$ and $l$ in the baseline model (i.e., the number of times tasks $k$ and $l$ execute during the response time of task $i$ in the baseline model)  \\
         \hline
         $R^\text{PAIR,0}_i$ & the converged worst-case response time of task $i$ in the post-violation state model without \acron interrupt latency. \\
         \hline
         $R^\text{PAIR}_i$ & the converged worst-case response time of task $i$ in the post-violation state model with \acron interrupt latency. \\
         \hline
         $C_{\text{PAIR}}$ & the interrupt path latency introduced by \acron, measured in Section~\ref{sec:performance-costs} \\
         \hline
         $n_p$ & simplified ceiling term of task $l$ in the post-violation model, shared by $R^\text{PAIR,0}_i$ and $R^\text{PAIR}_i$ (i.e., the number of times task $l$ executes within the response time of task $i$, given task $k$ only executes for $\mathit{budget}_k$ time) \\
         \hline
    \end{tabular}
    }
\end{table}

Because $R_i$ appears on both sides of the equation, this is a recursive (fixed-point) equation rather than a closed-form expression. The notation $R^{(n)}_i$ denotes $R_i$ at iteration $n$. Starting from $R^{(0)}_i = C_i + B_i$, the equation is iterated until it converges (i.e., when $R^{(n)}_i = R^{(n-1)}_i$) or the response time is deemed to exceed task $i$'s deadline. Upon convergence, $R^{(n)}_i$ becomes the fixed-point $R_i$. For all remaining equations, we assume $R_i$ has already converged and thus simplify $R^{(n)}_i = R^{(n-1)}_i = R_i$.

For the baseline analysis, we consider three benign tasks: the task under analysis $i$ and two higher-priority tasks, $k$ and $l$, both of which are assumed to execute according to their normal periodic behavior. The baseline response time of task $i$ in this scenario is consequently expressed in Equation~\ref{eq:baseline-resp-time}:

\begin{equation}\label{eq:baseline-resp-time}
R_i^{\mathrm{baseline}}
=
C_i + B_i
+
\left\lceil
\frac{R_i^{\mathrm{baseline}}}{T_k}
\right\rceil C_k
+
\left\lceil
\frac{R_i^{\mathrm{baseline}}}{T_l}
\right\rceil C_l
\end{equation}

In the baseline model, task $k$ is not exploited and runs as intended, providing a reference point for the next scenario. In the post-violation model next, task $k$ models the exploited task, which \acron detects and removes. We simplify Equation~\ref{eq:baseline-resp-time} using constants $n_k$ and $n_l$ to represent the ceiling of tasks $k$ and $l$, obtaining the baseline worst-case response time ($R^{\text{baseline}}_i$) of task $i$ in Equation~\ref{eq:baseline-resp-time-simpl}:

\begin{equation}\label{eq:baseline-resp-time-simpl}
R_i^{\mathrm{baseline}}
=
C_i + B_i
+
n_k C_k
+
n_l  C_l
\end{equation}

\subsection{Post-Violation State Model}

We next model the system state after \acron has removed a violating task.
In particular, task $k$ is assumed to execute for some bounded amount of time before a violation was detected, at which point it will be removed and prevented from any further execution by \acron.
Let $\mathit{budget}_k$ denote the amount of execution consumed by task $k$ before it was removed from \AR by \acron.
Since task $k$ can execute for any portion of its expected execution time, the model permits $\mathit{budget}_k$ to hold any value between 0 and $n_k C_k$.

Following revocation, task $k$ contributes no further periodic interference to the response-time of task $i$.
Its contribution is therefore represented as a one-shot term, yielding the response-time equation in Equation~\ref{eq:resp-time-case1}:

\begin{equation}\label{eq:resp-time-case1}
R_i^{\mathrm{PAIR,0}}
=
C_i + B_i
+
\mathit{budget}_k
+
\left\lceil
\frac{R_i^{\mathrm{PAIR,0}}}{T_l}
\right\rceil C_l
\end{equation}

In Equation~\ref{eq:resp-time-case1}, the superscript $\mathrm{PAIR,0}$ denotes the idealized \acron case with zero enforcement overhead. This formulation preserves the execution interference that has already occurred before revocation while preventing future execution of $k$.
%
%
%
However, \acron's interrupt path introduces a measurable delay (recall from Section~\ref{sec:performance-costs}) between the detection of a violation in task $k$ and the yield into a pending task. This behavior must be incorporated into the response-time model to determine whether the non-degradation property holds for the implemented system.
To account for this, we extend the response-time equation with a constant interrupt path latency, denoted by $C_{\mathrm{PAIR}}$. The resulting response-time equation is shown in Equation~\ref{eq:resp-time-case2}:

\begin{equation}\label{eq:resp-time-case2}
R_i^{\mathrm{PAIR}}
=
C_i + B_i
+
\mathit{budget}_k
+
C_{\mathrm{PAIR}}
+
\left\lceil
\frac{R_i^{\mathrm{PAIR}}}{T_l}
\right\rceil C_l
\end{equation}

This model captures both competing effects of \acron: the
removal of future interference from $k$ and the finite cost required to perform the \textit{kill-and-yield} operation.

The value of $C_{\mathrm{PAIR}}$ is obtained from measurements of the PAIR prototype in Section~\ref{sec:performance-costs}: between 134.7$\mu$s and 136.7$\mu$s.
Accordingly, we update the model to set $C_{\mathrm{PAIR}}$ equal to the rounded-up integer value of their mean: 136 $\mu$s.
Additionally, we simplify the expression for $R^{\text{PAIR}}_i$ by using $n_p$ to represent the ceiling factor in the post-violation model (e.g., the number of times task $l$ interferes with task $i$), obtaining the final expression in Equation~\ref{eq:resp-time-case2-simplified}:

\begin{equation}\label{eq:resp-time-case2-simplified}
R_i^{\mathrm{PAIR}}
=
C_i + B_i
+
\mathit{budget}_k
+
n_p C_l
+
136
\end{equation}

\subsection{Solving for Degradation Condition}\label{sec:constraint-solving}

We use the final simplified Equations~\ref{eq:baseline-resp-time-simpl} and~\ref{eq:resp-time-case2-simplified} to represent the response time of task $i$ under the two described scenarios.
We perform symbolic analysis to determine the condition under which the response time of task $i$ executing on an \acron-enabled system degrades compared to the baseline system. We implemented this using SymPy~\cite{sympy} in Python using 120 lines of code.

The variables $C_i$, $B_i$, $n_k$, $C_k$,  $C_l$, $n_l$, $budget_k$, and $n_p$ are defined as symbolic variables, while $C_{\text{PAIR}}$ is initialized to concrete integer value 136. 
$R^{\text{baseline}}_i$ from Equation~\ref{eq:baseline-resp-time-simpl} and $R^{\text{PAIR}}_i$ from Equation~\ref{eq:resp-time-case2-simplified} are defined as expressions using the corresponding symbolic variables. 

Our solver implemented using SymPy takes the difference of $R^{\text{PAIR}}_i$ and $R^{\text{baseline}}_i$ to generate a new expression corresponding to the non-degradation condition: $R^{\text{PAIR}}_i - R^{\text{baseline}}_i = 0$.
The evaluation of this expression returned the non-degradation condition in Equation~\ref{eq:non-degrad-cond}:

\begin{equation}\label{eq:non-degrad-cond}
    n_k C_k \geq budget_k + 136 + C_l (n_p-n_l).
\end{equation}

This expression states that \acron's interruption and removal of task $k$ does not degrade task $i$'s response time as long as task $k$'s baseline interference (e.g., its execution count $n_k$ multiplied by its execution time $C_k$) is at least as large as the combined cost of intervention:
\begin{enumerate}[nosep]
    \item the execution that task $k$ completed before being interrupted ($budget_k$); 
    \item \acron's interrupt path latency ($C_{\text{PAIR}} \approx$ 136 $\mu$s); 
    \item the added interference from task $l$ (the term $C_l (n_p-n_l)$) caused by task $l$ executing a different number of times once $k$ stops competing for processor time. 
\end{enumerate}

\input{figs/waveforms.tex}

When \acron does not affect the number of task $l$ executions (e.g., $n_p = n_l$), this constraint can simplify into the more intuitive expression shown in Equation~\ref{eq:non-degrad-cond-simpl}:

\begin{equation}\label{eq:non-degrad-cond-simpl}
    n_k C_k -  budget_k \geq 136.
\end{equation}

This scenario would occur when task $l$ has higher priority than task $k$, while task $k$ has higher priority than task $i$: since task $l$ already preempts task $k$ regardless of whether $k$ is revoked, removing $k$ does not change how often $l$ executes. 

In this case, degradation is determined purely by the $\mathit{budget}_k$
value. To analyze this case further, we assume that $\mathit{budget}_k$ takes the form $n_{bv} C_k$, where $n_{bv}$ represents the number of complete
executions of task $k$ that occur \textit{before a violation}. We use this substitution to obtain Equation~\ref{eq:solve-n_bv}:

\begin{equation}\label{eq:solve-n_bv}
    (n_k - n_{bv}) C_k \geq 136.
\end{equation}

This states s a direct relationship between \acron's enforcement
cost and the amount of execution eliminated by revoking task $k$. 
%
%
When $n_{bv} = n_k - \frac{136}{C_k}$, enforcement by \acron does not impact the response-time of task $i$. A lower $n_{bv}$ (i.e., violation by task $k$ detected earlier) will result in an improved response time for task $i$. A higher $n_{bv}$ (violation detected later) will incur an overhead bounded by the enforcement time duration. Note that any violation by a lower priority task has no effect on the response time of $i$.

\section{Case Study}\label{sec:case-study}

As a case study, we simulate the application example from Section~\ref{sec:application}, showing that \acron enables a system with a compromised task to remain online until other tasks complete execution. The simulated system includes four tasks:
\begin{itemize}[nosep=*]
    \item Task 1: A syringe pump motor and dosage control~\cite{opensyringe};
    \item Task 2: A sensor to monitor a patient's temperature~\cite{tempsensor};
    \item Task 3: A communication module using the openMSP430's peripheral UART interface;
    \item Task 4: A liquid crystal display~\cite{lcd_repo}.
\end{itemize}
Task 1 implements a stepper motor that advances the syringe plunger in micro-steps, with each pulse corresponding to one step. Based on the syringe dimensions and threaded rod pitch, the program computes approximately 13 micro-steps per milliliter (ml), yielding approximately 65 motor steps for a fixed 5 ml bolus. We modify this implementation with a counter, initialized to zero, that updates GPIO output port \texttt{p3\_out} with the latest pulse count; a functional execution therefore ends with \texttt{p3\_out} equal to 64 (\texttt{0x40}).

We evaluate this end-to-end example in two scenarios: (1) a benign execution and (2) an execution in which the communication module (task 3) is attacked via a simulated buffer overflow in input parsing, exploited to hijack the program's control flow. Waveforms for both appear in Figures~\ref{fig:case-study-benign} and~\ref{fig:case-study-attack}.

Both figures share four key signals.
\texttt{p3\_out} is a GPIO peripheral output showing task 1's pulse-step counter.
The remaining three signals capture \acron's run-time behavior. \texttt{AR\_en} is the bitmask representation of \AR stored internally to \acron hardware.
In this case study, \acron supports 16 concurrent tasks, so \texttt{AR\_en} is a 16-bit bitmask initialized to \texttt{0xffff} since all tasks initially belong to \AR. 
\texttt{cfi\_violation} is an input signal from \IM indicating a detected CFI violation. 
Finally, \texttt{zombify\_irq} is \acron's non-maskable interrupt output, which triggers the RTOS's \textit{kill-and-yield} functionality and simultaneously updates \texttt{AR\_en} when set.

The remaining waveform signals are monitoring signals added during simulation to identify which module is executing. \texttt{in\_task1}, \texttt{in\_task2}, \texttt{in\_task3}, and \texttt{in\_task4} are binary signals indicating when tasks 1 through 4 are executing, respectively, computed by comparing $PC$ to each task's static memory bounds in program memory (omitted from the waveforms for brevity). Similarly, $PC$ is compared to the RTOS region's bounds to produce \texttt{in\_RTOS}, signaling when the RTOS is executing.

Figure~\ref{fig:case-study-benign} shows the waveform for the benign execution. Task 3 runs first, receiving input from the network. Once a command is parsed, the RTOS schedules the remaining tasks, which then execute periodically with intermittent RTOS switching between them. No CFI violation is detected during this execution, so \IM never sets \texttt{cfi\_violation}, \acron's \texttt{zombify\_irq} remains cleared throughout, and \texttt{AR\_en} stays at its initialized value \texttt{0xffff}, indicating no task was removed from \AR. The syringe control task completes at $\approx$24.62 ms, having executed all dosage pulses, as reflected by \texttt{p3\_out} reaching \texttt{0x40} at the end of its execution. The full system completes at $\approx$26.45 ms.

Figure~\ref{fig:case-study-attack} showcases the waveform after simulating the attacked execution. 
In this scenario, a CFI violation is detected during the execution of the communication module (task 3). This occurs at $\approx$10.71 ms, as shown with the rising edge of \texttt{cfi\_violation} at this time. This corresponds to \acron setting \texttt{zombify\_irq} to trigger the removal of task 3 via the RTOS's \textit{kill-and-yield} functionality. At this same time, \acron updates \AR to remove the violating task. This is shown in the transition of \texttt{AR\_en} from \texttt{0xffff} to \texttt{0xfffb}. 

At this point, task 3 is no longer authorized to execute. Compared to the benign execution in Figure~\ref{fig:case-study-benign}, the second pulse of \texttt{in\_task3} is much shorter, and it does not continue after the \texttt{cfi\_violation} is set. This portion of the waveform shows that task 3 is prevented from further execution. However, similarly to the benign case shown in Figure~\ref{fig:case-study-benign}, the RTOS follows the same schedule and enters back into task 1, shown with \texttt{in\_task1} being the next task monitor signal to be set after the second pulse of \texttt{in\_task3}. 

The three remaining non-violating tasks (task 1, task 2, and task 4) continue executing until all have completed their operation. In this scenario, due to the removal of task 3 midway through its execution, the entire system completes execution early. In this case, all tasks finish executing at $\approx$25.31 ms compared to $\approx$26.45 ms in the benign execution ($\approx$ 1.14 ms early). Similarly, task 1 finishes early compared to the benign execution: in this scenario, it completes at $\approx$23.47 ms compared to $\approx$24.62 ms in the previous execution. 

This case study demonstrates \acron's non-degradation property in a realistic, multi-task embedded system rather than just analytically. When task 3 is compromised and revoked mid-execution, the remaining tasks are unaffected in their functional correctness, such as the syringe pump control task completing all of its expected pulses. Rather than degrading system performance, revocation actually allows the system to finish earlier since the compromised task no longer competes for CPU time. This confirms that \acron can remove a violating task at run time without disrupting the timing or correctness of the tasks that continue running. 

\section{Related Work}\label{sec:related-work}

\textbf{CFI in Real-Time Systems.} Many works have explored CFI in embedded and real-time systems~\cite{cfi_rts_survey}. InsectACIDE~\cite{insectacide} proposes a method for CFI for ARM Cortex-M MCUs leveraging CoreSight tracing extensions to record control flow transfers without incurring run-time overhead. Then, recorded control flow transfers are checked for integrity violations during idle times. 
RECFISH~\cite{recfish} demonstrates an RTOS-compatible CFI method for ARM Cortex-R MCUs through shadow stacks and function-label checks via software instrumentation. %
FastCFI~\cite{fastcfi} also uses ARM extensions, but instead uses an FPGA to store and check a CFG. 
Regardless of their specificities, \acron differs from each in that it accounts for the post-violation phase, allowing non-violating tasks to continue their execution. 

EILID~\cite{eilid} proposes a hybrid method for low-end MCUs and evaluates atop openMSP430. It uses code instrumentation for trampolines that jump to a hardware-protected CFI monitor. 
\acron+\IM incurs more hardware overhead, while EILID incurs more run-time overhead and an increase in code size.
Similar to \acron, EILID triggers an interrupt upon violations and also enables software updates. However, EILID's interrupt jumps directly to the software update function.
\acron's interrupt instead disables a violating task before returning to the next high-priority benign task.
Due to parallel goals, \acron and EILID could coexist, using \acron to provide post-violation availability/integrity with EILID as a low-cost \IM. 

\textbf{Availability-focused mechanisms.} Active roots of trust (ARoTs)~\cite{acfa,traces,garota,aion,sdrad,caramel} ensure a trusted function is guaranteed to execute upon a corresponding hardware signal or upon MCU reset.
The concept of ARoTs was proposed in GAROTA~\cite{garota}, a formally verified architecture that provides two properties regarding the execution of a trusted interrupt service routine (ISR): (1) the guaranteed triggering of the trusted ISR upon its interrupt source (e.g., GPIO, timer), and (2) the guaranteed re-triggering of the trusted ISR upon boot or reset. 
Several works built upon GAROTA to realize ARoTs for various purposes~\cite{acfa,caramel,traces}.

AION~\cite{aion} is an ARoT that builds upon Sancus TEE~\cite{Sancus17} to provide strong availability guarantees to a trusted scheduler in the presence of compromised software.
Like AION, we also built our prototype using openMSP430 as an MCU and RIOT as an RTOS.
AION ensures the availability of a trusted Sancus-aware scheduler by incorporating hardware-based atomicity monitor and exception engine modules. 
While both AION and \acron invoke the RTOS scheduler upon a violation signal, they pursue complementary objectives.
AION aims to preserve the availability of a TEE-protected trusted scheduler to fairly allocate execution time to tasks, whereas \acron focuses on preventing tasks that violate run-time integrity policies from executing until they are updated, thereby preserving availability to other non-violating tasks on the same platform. This is provided even if other software, including the scheduler, attempts to re-enter it. 
Given the orthogonal goals, \acron and AION combined could provide a strong security service: AION's trusted scheduler with \acron's enforcement against re-entering of violating tasks could return both fair scheduling and post-violation containment.

SDRaD~\cite{sdrad} leverages Intel PKU to compartmentalize applications and to trigger rollback of execution state upon a violation. SDRaD is tailored towards restoring prior execution context with custom handlers to avoid replay of the attack and targets rich application software on COTS processors, whereas PAIR targets resource-constrained MCUs operating under hard real-time constraints, enforcing post-violation task containment entirely in hardware, and explicit integration with RTOS scheduling semantics.

\textbf{Formally verified MCU mechanisms.} Several prior works share this work's strategy of formal modeling/verification via LTL. VRASED~\cite{vrased} is a formally verified hybrid remote attestation architecture, using formally verified cryptographic libraries to compute attestation evidence and LTL to model and verify the hardware extension that monitors and protects the attestation software. APEX~\cite{apex} and ASAP~\cite{asap} propose formally verified architectures for \textit{proofs of execution}, demonstrating that a given software executed completely and without modification, and that reported outputs were generated by that same execution. UCCA~\cite{ucca} is a formally verified architecture providing isolation of untrusted code, with formal verification ensuring its hardware extension provide return integrity, stack integrity, and critical memory region protection.

\section{Discussion \& Future Work}\label{sec:discussion}

\textbf{Integrity Monitor Soundness.} \acron's guarantees are conditioned on the correctness of the \IM, as \acron takes \textit{$violation_{IM}$} as input from \IM. We do not verify the correctness of the \IM in use. A false negative in the \IM leaves \acron with nothing to act on, silently forfeiting desired properties despite \acron behaving correctly. Conversely, a false positive would revoke a benign task, directly working against \acron's availability goal. As a result, \acron's security is only as strong as the soundness of the paired \IM. As there have been proposals for formally-verified CFI mechanisms~\cite{kcofi,ceesay2024mucfi,lalande2014software}, pairing \acron with such an \IM would ensure end-to-end formal guarantees.

\textbf{Interdependent Tasks.} \acron assumes tasks are functionally independent. It is possible that an application is designed such that its tasks coordinate via locks, mutexes, or semaphores. As such, removing a violating task from \AR mid-execution may leave shared locks unreleased, causing deadlock. Addressing this requires further consideration of RTOS locking primitives, including monitoring of its access and ability to unlock in special cases from \acron's \sw protected region. We leave this as a direction for future work.

\textbf{Side-channel attacks.} \acron's software update routine authenticates update requests using a secret key. A naive implementation may be vulnerable to timing side-channel attacks if it enables any secret-dependent execution paths. This can be mitigated by ensuring the authentication routine is implemented as constant-time code. Cache-timing side-channels are a common concern for cryptographic routines on general-purpose processors. However, this is not applicable to devices in \acron's scope, which often do not incorporate caches. However, this is an important consideration when extending to more complex processors in future work.

\textbf{Multi-Core Embedded Systems.} Extending \acron to multi-core embedded systems introduces challenges. For example, \AR state must be maintained atomically across concurrent contexts, and kill-and-yield semantics must account for cross-core scheduling. An extension of PAIR to support multi-core is an intriguing direction for future work.

\textbf{Task Criticality.} \acron provides an effective balance for scenarios in which preventing a task from continuing is safer than allowing it to continue in an unexpected state, regardless of the task's criticality. Although applicable to many scenarios (as described in Section~\ref{sec:application}), there are scenarios in which preventing a compromised high-criticality task from continuing may still be unideal (e.g., automotive vehicles). 
Future work could explore enhancing PAIR \sw to differentiate violation handling based on criticality: lower-criticality tasks are immediately removed from \AR, while higher-criticality tasks are re-executed once before removal, leveraging aspects of PAIR and related works~\cite{aion,sdrad,garota}.    

\section{Conclusion}
In this work, we propose \acron to preserve task availability and system-wide run-time integrity. By ensuring that only the offending task is prevented from executing, \acron enables non-offending critical real-time tasks to continue executing. We designed, implemented, and evaluated a formally verified and open-source prototype of \acron\footnote{The source code will be made available after peer review.}, finding that its design has reasonable costs, incurring no runtime overhead while imposing minimal hardware and performance overheads.

\bibliographystyle{plain}
\bibliography{references}

\end{document}

%% file: commands_packages.tex
\usepackage{usenix}

\usepackage{amsmath}

\usepackage{amssymb,amsfonts}
\usepackage{subcaption}
\usepackage{algorithmic}
\usepackage{graphicx}
\usepackage{textcomp}
\usepackage{standalone}
\usepackage{xcolor}
\def\BibTeX{{\rm B\kern-.05em{\sc i\kern-.025em b}\kern-.08emT\kern-.1667em\lower.7ex\hbox{E}\kern-.125emX}}
\usepackage[utf8]{inputenc}
\usepackage{booktabs} 
\usepackage{graphicx}
\usepackage{units}
\usepackage{blindtext}
\usepackage{multirow}
\usepackage{xcolor,xspace}
\usepackage{amsmath}
\usepackage{paralist}
\usepackage{mdframed}
\usepackage{pict2e}
\usepackage{listings}
\usepackage{color}
\usepackage{threeparttable}
\usepackage{array}
\usepackage{booktabs}
\usepackage{pifont}
\usepackage{url}
\usepackage{comment}
\usepackage{amsmath}
\usepackage{lipsum}
\usepackage{arydshln}
\usepackage{flushend}
\usepackage{latexsym}
\usepackage{enumitem}
\usepackage{fancyhdr}
\usepackage[english]{babel}
\usepackage{tabularx}
\input{glyphtounicode}

\usepackage[ruled]{algorithm2e}
\usepackage{algorithmic}
\usepackage{url}
\usepackage[
    n,
    advantage,
    operators,
    sets,
    adversary,
    landau,
    probability,
    notions,
    logic,
    ff,
    mm,
    primitives,
    events,
    complexity,
    asymptotics,
    keys]{cryptocode}

\usepackage{tikz}                    
\usepackage{pgfplots}
\usetikzlibrary{arrows, patterns, automata, positioning}
\pgfplotsset{compat=1.18}
\usepgfplotslibrary{groupplots}
\usepackage{circuitikz}
\usepackage{tikz-timing}
\newcommand{\ignore}[1]{}

\newcommand{\acron}{\textsf{{PAIR}}\xspace}
\newcommand{\AR}{\textsf{AR}\xspace}
\newcommand{\IM}{\textsf{IM}\xspace}
\newcommand{\TR}{\textsf{TR}\xspace}
\newcommand{\sw}{\textsf{SW}\xspace}

\renewcommand\adv{\ensuremath{\sf{\mathcal Adv}}\xspace}

\newlist{myenumerate}{enumerate}{1}
\setlist[myenumerate]{
  label=(\arabic*),
  leftmargin=12pt,
  nosep,
}

\newlist{myitemize}{itemize}{1}
\setlist[myitemize]{
    label=$\bullet$,
    leftmargin=12pt,
    nosep,
}

\AtBeginDocument{%
  \providecommand\BibTeX{{%
    Bib\TeX}}}

\usepackage{appendix}

%% file: ltl_specs/goals_ltl.tex
\begin{figure}[t]
\small
\fbox{
    \parbox{\columnwidth}{
        
        \textbf{\underline{Definition 1:}} Post-Violation Availability 
        \begin{equation*}\label{def:goal1}
            G :\{violation_{\acron} \lor violation_{IM} \rightarrow X(trigger)) \}
        \end{equation*}
        \begin{equation*}
            G: \{(PC \in TR) \land (task_{id} = i) \land trigger \rightarrow \neg X(T_{i} \in AR)\}
        \end{equation*}

        \textbf{\underline{Definition 2:}} Post-Violation Integrity
        \begin{equation*}\label{def:goal2}
            G :\{(PC \in TR) \land (task_{id} = i)  \rightarrow (T_{i} \in AR) \lor X(trigger)\}
        \end{equation*}
        }
}
\vspace{-1em}
\caption{Formal Specification of security properties}
\label{fig:formal-goals}
\end{figure}

%% file: ltl_specs/task_monitor_ltl.tex

\begin{figure*}[t]
\fbox{
\small
    \parbox{\textwidth}{
        \textbf{\underline{Axiom:} Tracked Task Execution:}
        \begin{equation}\label{ltl:track-task}
            G: \{PC \leq T^{max}_{i} \land PC \geq T^{min}_{i} \rightarrow (task_{id} = i) \} \forall i \in \{0, N-1\}
        \end{equation}
    
        
    

        
        \textbf{Task Memory Access Integrity:}
        \begin{equation}\label{ltl:task-mon-violation-2}
            \begin{aligned}
            G : \{(R_{en} \lor &W_{en}) \land (D_{addr} \in D_{\acron}) \land  \neg(PC \in SW) \rightarrow violation_{\acron})\}
            \end{aligned} 
        \end{equation}
        \begin{equation}\label{ltl:task-mon-violation-3}
            \begin{aligned}
                G: \{ W_{en} \land &(D_{addr} \in PMEM) \land \neg(PC \in SW) \rightarrow violation_{\acron} \}
            \end{aligned}
        \end{equation}
        \begin{equation}\label{ltl:pair-violation}
            \begin{aligned}
                G: (violation_{\acron} \rightarrow X(trigger))
            \end{aligned}
        \end{equation}
    }
}
\vspace{-1em}
\caption{Sub-properties to monitor task execution.}
\label{fig:task-mon}
\end{figure*}

%% file: ltl_specs/avail_mon_ltl.tex
\begin{figure*}[t]
\small
\fbox{
    \parbox{\textwidth}{
        \textbf{\underline{Definitions}}: $N$-bit bitmasks.
        \begin{equation*}
            \texttt{mask}(id) := (1 << id), zero := 0_N, set := 1_N
        \end{equation*}
        




        \textbf{Triggering upon integrity monitor violation:}
        \begin{equation}\label{ltl:generate-irq}
            \begin{aligned}
            G: \{violation_{IM} \rightarrow X(trigger)\}
            \end{aligned}
        \end{equation}
        \begin{equation}\label{ltl:clear-irq}
            \begin{aligned}
            G: \{trigger \land irq \rightarrow \neg X(trigger)\}
            \end{aligned}
        \end{equation}

        \textbf{Reducing \AR coverage upon trigger:}
        \begin{equation}\label{ltl:update-AR-trigger}
            \begin{aligned}
            G: \{((PC \in TR) \land (task_{id} = i) ) \land trigger \rightarrow (X(AR_{en}) \And \texttt{mask}(task_{id})) = zero)\}
            \end{aligned}
        \end{equation}

        \textbf{Triggering upon re-execution of violating task:}
        \begin{equation}\label{ltl:trigger-on-reenter}
            \begin{aligned}
            G: \{((PC \in TR) \land (task_{id} = i) ) \land (AR_{en} \And \texttt{mask}(task_{id})) = zero) \rightarrow X(trigger)\}
            \end{aligned}
        \end{equation} 
        
        \textbf{Resetting $AR$ after software update:}
        \begin{equation}\label{ltl:reset-AR}
            \begin{aligned}
            G : \{PC = SW_{exit} \rightarrow X(AR_{en}) = set\}
            \end{aligned}
        \end{equation}

    }
}
\vspace{-1em}
\caption{Sub-properties to maintain \AR.}
\label{fig:aviail-mon-spec}
\end{figure*}

%% file: ltl_specs/proofs.tex
\begin{figure}[t]
\small
\fbox{
    \parbox{0.95\columnwidth}{
        \textbf{\underline{Theorem 1}:}
        \vspace{-1em}
        \begin{center}
            LTLs~\ref{ltl:task-mon-violation-2},~\ref{ltl:task-mon-violation-3},~\ref{ltl:pair-violation},~\ref{ltl:generate-irq}-~\ref{ltl:clear-irq} $\rightarrow$ Definition 1
        \end{center}

        \textbf{\underline{Theorem 2}:}
        \vspace{-1em}
        \begin{center}
            LTLs~\ref{ltl:track-task},~\ref{ltl:update-AR-trigger},~\ref{ltl:trigger-on-reenter},~\ref{ltl:reset-AR} $\rightarrow$ Definition 2
        \end{center}

    }
}
\vspace{-1em}
\caption{Theorems that \textbf{[P1]} and \textbf{[P2]} are upheld.}
\label{fig:proofs}
\end{figure}

%% file: figs/fsm-trigger.tex
\begin{figure}[t]
\begin{center}

\noindent\resizebox{\linewidth}{!}{%
	\begin{tikzpicture}[->,>=stealth',auto,node distance=8.0cm,semithick]
		\tikzstyle{every state}=[minimum size=1.5cm]
		\tikzstyle{every node}=[font=\scriptsize]

		\node[state] (A) {\textit{Exec}};
		\node[state, fill={rgb:black,1;white,3}] (B) [right=2cm of A,align=center]	{\textit{Trigger}};

		\path[->,every loop/.style={looseness=8}] 
			(A) edge [loop left, min distance=4mm] node {\textit{else}} (A)
			(B) edge [loop right, min distance=4mm] node {\textit{else}} (B);
  		
        \draw[->] (A.290) -- node[rotate=0,below, align=center,auto=right] {\shortstack{
        $[W_{wen} \land data_{addr} \in PMEM \land \neg(PC \in SW)] \lor$ \\
        $[(R_{en} \lor W_{wen}) \land data_{addr} \in D_{\acron} \land \neg(PC \in SW)] \lor$ \\
        $[violation_{IM} \land \neg reset] \lor [(PC \in T_{i}) \land \neg (T_{i} \in AR)]$
        }
        } (B.250);
        
        \draw[<-] (A.70) -- node[rotate=0,above] {\shortstack{
        $(trigger \land irq) \lor reset$}
        }
        (B.110);
	\end{tikzpicture}
}
\end{center}
\vspace{-1.5em}
\caption{Verified FSM for LTLs~\ref{ltl:task-mon-violation-2},~\ref{ltl:task-mon-violation-3},~\ref{ltl:generate-irq},~\ref{ltl:clear-irq}.}
\label{fig:fsm-trigger}
\end{figure}

%% file: figs/fsm-ar.tex
		
        
        
                






\begin{figure}[t]
\begin{center}
\resizebox{0.8\columnwidth}{!}{%
	\begin{tikzpicture}[->,>=stealth',auto,node distance=8.0cm,semithick]
		\tikzstyle{every state}=[minimum size=2cm]
		\tikzstyle{every node}=[font=\large]
		\node[state] (0,0) (A) {\shortstack{\textit{Monitor}}};
        \node[state] [right=2cm of A,fill={rgb:black,1;white,3}] (B){\shortstack{\textit{Revoke}}};
        \node[state] (C) [left=2cm of A,fill={rgb:black,1;white,9}] {\shortstack{\textit{Reinstate}}};
                
		\path[->,every loop/.style={min distance=1mm, looseness=8}] 
			(A) edge [loop above, min distance=4mm] node {\textit{else}} (A);

        \draw[->] (A.345) -- node[rotate=0,below, align=center,auto=right]{\shortstack{$trigger$}} (B.195);

        \draw[<-] (A.15) -- node[rotate=0,above] { }(B.165);

        \draw[<-] (C.345) -- node[rotate=0,below, align=center,auto=right]{\shortstack{$PC = SW_{exit}$}} (A.195);

        \draw[<-] (A.165) -- node[below] {} (C.15);

	\end{tikzpicture}
}
\caption{Verified FSM for LTLs
~\ref{ltl:update-AR-trigger},~\ref{ltl:reset-AR}.
}
\label{fig:fsm-ar}
\end{center}
\end{figure}

%% file: figs/hw-cost.tex
\newcommand{\LUTcolor}{black!50}
\newcommand{\FFcolor}{black!10}
\newcommand{\xorigin}{1}

\begin{figure}[t]
    \centering
    \resizebox{\columnwidth}{!}{
        \begin{tikzpicture}
            \begin{axis}[
                width=1.25*\columnwidth,
                height=4.5cm,
                ymin=0, ymax=1600,
                ytick={0,750,1500}, 
                symbolic x coords={A,B,C,D,E,F,G},
                enlarge x limits = .1,
                xtick=data,
                ylabel={Total},
                xticklabels={
                    \footnotesize{\acron}, 
                    \footnotesize{GAROTA}, 
                    \footnotesize{ACFA}, 
                    \footnotesize{VRASED}, 
                    \footnotesize{UCCA}, 
                    \shortstack{\footnotesize EILID \\ \footnotesize(CASU)},
                    \shortstack{\footnotesize AION \\ \footnotesize(Sancus)}
                },
                ybar=0.1pt,
                scaled y ticks =false,
                bar width=12pt,
                legend style={at={(0.185,0.95)}}
            ]
            \addplot+[ybar, fill=\LUTcolor, draw=black, line width=0.25pt] coordinates {
            (A, 15)
            (B, 99)
            (C, 275)
            (D, 122)
            (E, 191)
            (F, 99)
            (G, 1464)
            };
        
            \addplot+[ybar, fill=\FFcolor, draw=black, line width=0.25pt] coordinates {
            (A, 16)
            (B, 33)
            (C, 202)
            (D, 37)
            (E, 265)
            (F, 34)
            (G, 814)
            };
            \legend{LUTs, FFs}
            \end{axis}    
        \end{tikzpicture}
    }
    \vspace{-2em}
    \caption{Hardware cost of \acron is negligible and significantly lower than related works.}
    \label{fig:hw-cost}
\end{figure}


%% file: figs/memory_usage.tex
\begin{table}[t]
    \centering
    \caption{Memory usage for applications with multiple tasks (MT), recursion (R), indirect forward branches (IF), and high/very high MCU suitability (SUI)~\cite{beebs_repo}.}
    \vspace{-1em}
    \resizebox{\columnwidth}{!}{
    \begin{tabular}{|l|c|c|c|}
         \hline
          &  & \multicolumn{2}{c|}{\textbf{Usage in bytes}}\\
         \textbf{Application} & \textbf{Attribute} & \multicolumn{1}{c}{\textbf{\acron}} & \multicolumn{1}{c|}{\textbf{\acron + IM}}\\
         \hline
         \texttt{sched\_round\_robin} & MT   & 12 (+0.073\%) & 24 (+0.146\%)\\
         \texttt{ipc\_pingpoing} & MT        & 12 (+0.073\%) & 14 (+0.085\%)\\
         \texttt{thread\_priority\_inversion} & MT & 16 (+0.097\%) & 22 (+0.134\%) \\
         \texttt{mutex\_sleep} & MT      & 8 (+0.048\%) & 12 (+0.073\%)\\
         \texttt{sched\_change\_priority} & MT & 8 (+0.048\%) & 12 (+0.073\%)\\
         \hline
         \texttt{crc\_32} & SUI              & 4 (+0.024\%)  &  8 (+0.048\%)\\
         \texttt{matrixmult} & SUI            & 4 (+0.024\%) &  12 (+0.073\%)\\
         \texttt{lcdnum} & IF, SUI     & 4 (+0.024\%) &  42 (+0.256\%)\\
         \texttt{cover} & IF, SUI      & 4 (+0.024\%) &  374 (+2.28\%) \\
         \texttt{cubic} & SUI      & 4 (+0.024\%) & 10 (+0.061\%) \\
         \texttt{fdct} & SUI                 & 4 (+0.024\%) & 10 (+0.061\%) \\
         \texttt{tarai} & R               & 4 (+0.024\%) &  18 (+0.109\%) \\
         \texttt{recursion} & R            & 4 (+0.024\%) &  26 (+0.159\%) \\
         \hline
    \end{tabular}
    }
    \label{tab:memory_usage}
\end{table}

%% file: figs/irq-latency.tex
\begin{figure}[t]
    \centering
        \begin{tikzpicture}
        \begin{groupplot}[
            group style={
                group name=myplots,
                group size=1 by 2,
                vertical sep=8pt, 
            },
            ybar,
            symbolic x coords={A,B,C,D},
            xtick=data,
            width=7cm,
            ymin=0,
        ]
    
    \nextgroupplot[
        ylabel={\footnotesize Mean Latency ($\mu$s)}, 
        ylabel style={xshift=-0.25cm},
        height=3cm,
        width=\columnwidth,
        ymin=128, 
        ymax=140,
        axis x line*=top,
        ytick={130, 135, 140},
        axis y line*=left,
        bar width=16pt,
        xticklabels={}, 
    ]
    \addplot+[fill=\FFcolor, draw=black, line width=0.25pt, error bars/.cd, y dir=plus,y explicit,error bar style={black}]
            coordinates {
            (A, 133.2)+- (0,0.346)
            (B, 136.7)+- (0,0.325)
            (C, 131.0)+- (0,0.325)
            (D, 134.7)+- (0,0.305)
            };
    
    \nextgroupplot[
        height=2cm,
        width=\columnwidth,
        ymin=0, 
        ymax=5, 
        ytick={0, 5},
        axis x line*=bottom,
        axis y line*=left,
        bar width=16pt,   
        xticklabels={
            \footnotesize{Call (H$\rightarrow$L)}, 
            \footnotesize{\acron (H$\rightarrow$L)}, 
            \footnotesize{Call (L$\rightarrow$H)}, 
            \footnotesize{\acron (L$\rightarrow$H)}, 
        },
    ]
    \addplot+[fill=\FFcolor, draw=black, line width=0.25pt, error bars/.cd, y dir=plus,y explicit,error bar style={red}]
            coordinates {
            (A, 133.2)+- (0,0.346)
            (B, 136.7)+- (0,0.125)
            (C, 131.0)+- (0,0.105)
            (D, 134.7)+- (0,0.305)
            };
    
    \end{groupplot}
    

    \draw[black, dotted, thick] (myplots c1r1.south west) -- ++(myplots c1r2.north west);

    \draw[black, dotted, thick] (.58,-0.24) -- (.58,.018);
    \draw[black, dotted, thick] (.58+1.9,-0.24) -- (.58+1.9,.018);
    \draw[black, dotted, thick] (.58+3.825,-0.24) -- (.58+3.825,.018);
    \draw[black, dotted, thick] (.58+5.75,-0.24) -- (.58+5.75,.018);
    
    \draw[black] (myplots c1r1.north east) -- (myplots c1r2.south east);
    \end{tikzpicture}
    \vspace{-2em}
    \caption{Mean \textit{kill-and-yield} latency of \acron interrupt by priority direction (high-to-low: H$\rightarrow$L, vs. low-to-high: L$\rightarrow$H) compared to a native call}
    \label{fig:kill-and-yield-time}
\end{figure}


%% file: figs/waveforms.tex
\begin{figure*}[t]
    \centering
    \includegraphics[width=\textwidth]{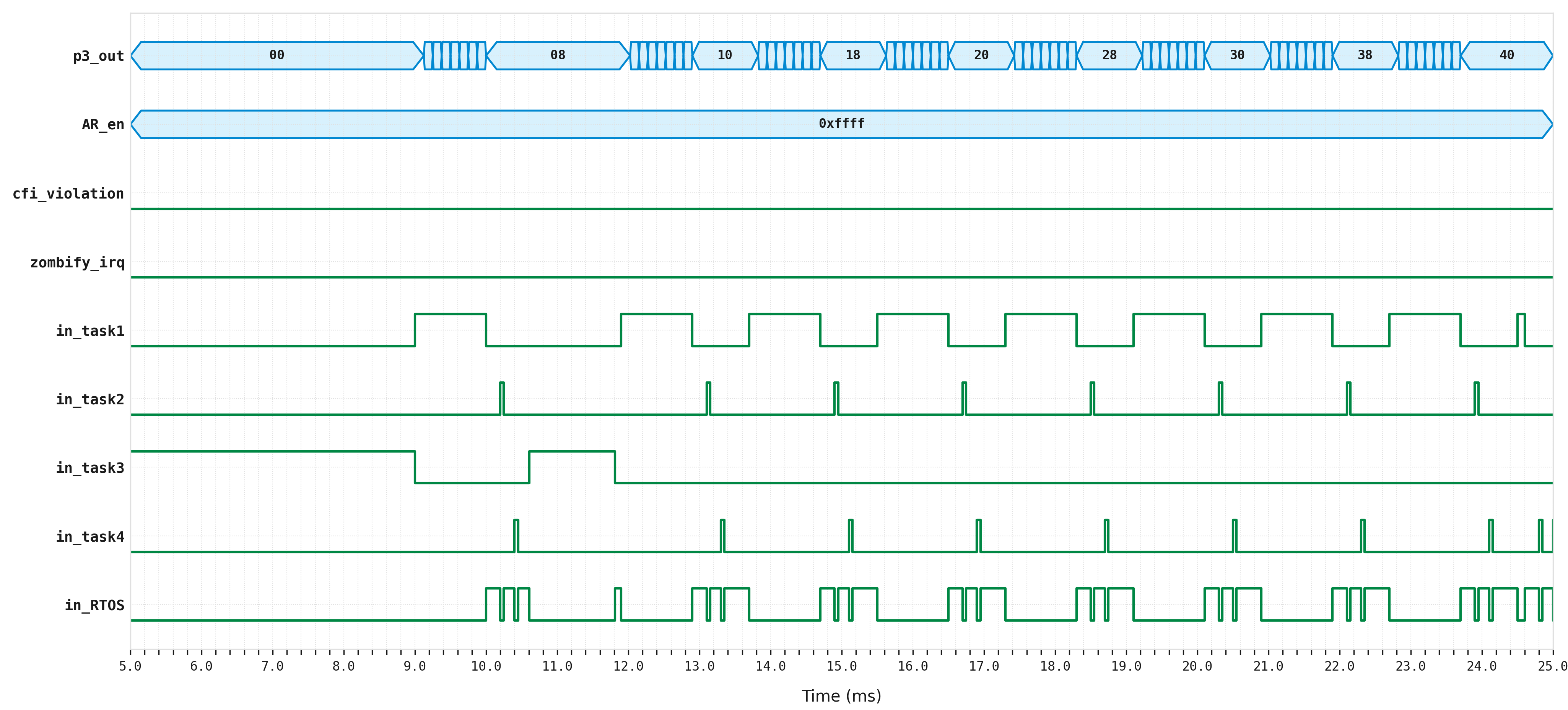}
    \caption{Case study: this waveform shows the result of a benign execution. During this execution, no \texttt{cfi\_violation} occurs, and thus \acron does not trigger its \texttt{zombify\_irq} signal. As a result, the syringe pump control task (\texttt{task\_1}) outputs its dosage in 64 steps (shown through its output on \texttt{p3\_out} showing \texttt{0x40}). All tasks complete by $\approx$26.45ms, with the syringe pump control completing by $\approx$24.62 ms.}
    \label{fig:case-study-benign}
    \end{figure*}
\begin{figure*}[t]
    \centering
    \includegraphics[width=\textwidth]{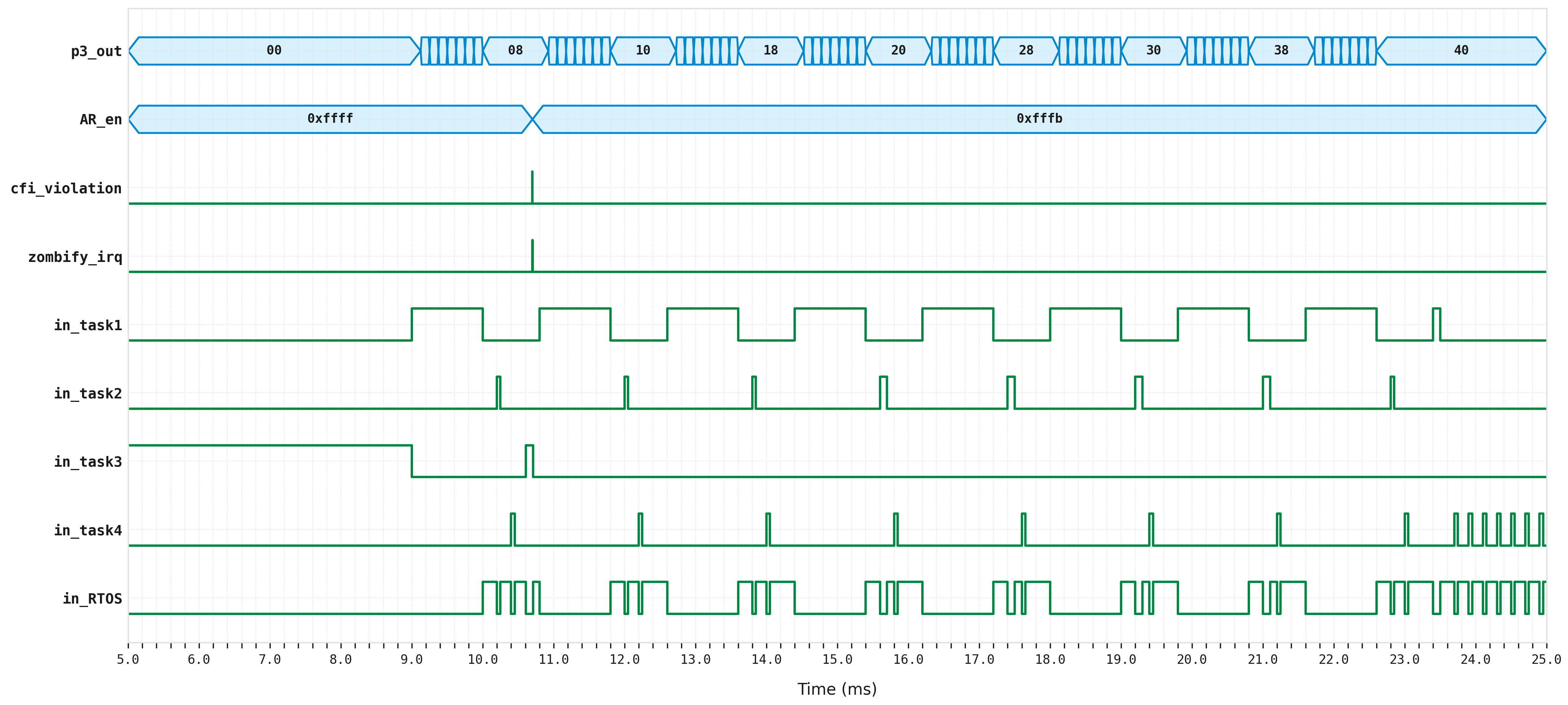}
    \caption{Case Study: this waveform shows the end result of an attacked execution, in which Task 3 (the communication module) has a CFI violation. At $\approx$10.71ms, the \texttt{cfi\_violation} signal is raised during the execution of Task 3. This results in the removal of Task 3 from AR (shown by transition of \texttt{AR\_en}=\texttt{0xffff} to \texttt{0xfffb}) and a trigger of \texttt{zombify\_irq}, \acron's non-maskable interrupt. By the end of the execution, all remaining tasks have completed in 25.31 ms, and the syringe dosage control completes at $\approx$23.47ms with all dosages given (e.g., with \texttt{p3\_out}=\texttt{0x40}, the expected value from the benign execution).}
    \label{fig:case-study-attack}
\end{figure*}